\documentclass[twoside,reqno]{HERON}
\usepackage{epsfig,cite,colordvi}
\usepackage{graphicx}
\usepackage{url}
\usepackage{amssymb,amsmath,amscd,epsf}
\usepackage{times}
\usepackage{makeidx}
\begin{document}

\title{Beyond the 2-body Interaction Paradigm: \\ 
\small The Case for Extended A-body Pairing Interaction in Nuclei}

\runningheads{Beyond the 2-body Interaction Paradigm}
{V.G. Gueorguiev}

\begin{start}

\author{Vesselin G. Gueorguiev}{1,2}

\index{Gueorguiev, V.G.} 

\address{Ronin Institute, Montclair, NJ, USA}{1}
\address{Institute for Advanced Physical Studies, Sofia, Bulgaria}{2}

\begin{Abstract}
We discuss modeling of nuclear structure beyond the 2-body interaction paradigm. Our first example is related to the need of three nucleon contact interaction terms suggested by chiral perturbation theory. The relationship of the two low-energy effective coupling parameters for the relevant three nucleon contact interaction terms  $c_D$ and $c_E$ that reproduce the binding energy of $^3$H and $^3$He has been emphasized and the physically relevant parameter region has been illustrated using the binding energy of $^4$He. Further justification of  A-body interaction terms is outlined based on the  Okubo-Lee-Suzuki effective interaction method used in solving the nuclear many-body problem within a finite model space. The third example we use is an exactly solvable A-body extended pairing interaction applied to heavy nuclei with a long isotopic chains; in particular using $^{132}$Sn and $^{208}$Pb as closed core system illustrates a remarkable relationship between the  extended pairing strength  $G(A)$ and the size of the valence space $\dim(A)$ for the members of these two isotope chain: $G(A)=\alpha\dim(A)^{-\beta}$ with $\alpha=259.436$ for Sn and $\alpha=366.77$ for Pb while the parameter $\beta$ is practically 1. These  three cases present evidence for the need of better understanding of the three-nucleon (NNN), four-nucleon (NNNN), and A-body interactions in nuclei either derived from ChPT or from a phenomenological considerations. 
\end{Abstract}
\end{start}

The high precision, QCD derived, nucleon interaction that describes the NN-scattering phase shifts and the deuteron, 
when applied to the light s- and p-shell nuclei points to the necessity of NNN-interaction terms\cite{Machleidt, Petr}. 
Thus, the conventional two-body interaction paradigm is challenged and the need of 3-body and possibly A-body interaction define a new research frontier.
The structure of the three-body terms has been studied previously using the meson exchange theory\cite{TM79}. 
However, with the advance of the Chiral Perturbation Theory (ChPT)\cite{TM'99,TM'01} the structure of the three-body interaction has been clearly identified and well justified via QCD. 

Higher many-body interaction terms (e.g. NNNN-interaction terms) are also part of the interaction as derived from QCD via ChPT\cite{Epelbaum}. 
The Okubo-Lee-Suzuki (OLS) effective interaction method, employed in solving the nuclear many-body theory, 
also introduces interaction terms beyond the common 2-body interaction\cite{Okubo, Lee-Suzuki}. 
All this seems to be pointing to the need of A-body interactions for the description of the nuclear structure. 
It also raises the question about the importance of the A-body interactions in very heavy nuclei. 
Fortunately, there is an exactly solvable A-body model -\textit{ the extended pairing model} - that is applicable as an A-body interaction to very heavy nuclei; 
therefore, it can help to address this question\cite{Feng,VGG2004,EPJ05}.

In the next section we briefly discuss the microscopic nuclear physics hamiltonian; 
the types of the high-precision NN-interaction potentials and their failure to properly account for the structure of the nuclei with more than two nucleons. 
In Sec. 3 we discuss the values of the $c_D$ and $c_E$ parameters of the NNN-intercation\cite{Petr} and their physically acceptable regions
as deduced from the binding energy of $^3$H, $^3$He, and $^4$He. 
In Sec. 4 we further extend our argument for A-body nuclear interactions by using the modern OLS effective interaction in finite model space method. 
In Sec. 5 we briefly discuss the results of applying the A-body Extended Pairing Interaction (EPI) to few long isotope chains like Sn and Pb nuclei. 
Last section is our conclusion about the needs of the future nuclear structure modeling based on A-body nuclear interactions.

\section{Modeling the Nuclear Interactions}

Unlike the electromagnetic and the gravitational interaction, the mathematical form of the nuclear interaction has been very elusive. 
This is due to the fact that the nuclear interaction arises non-trivially from the quark structure of the nucleons and thus related to the theory of the QCD. 
Never the less, the field of nuclear structure modeling has advanced significantly, based on general quantum mechanical principals and techniques. 
In particular, the microscopic approach has been very successful especially with the advance of computational techniques and computer power 
that have allowed for the construction of effective high-precision meson and/or QCD derived NN-potentials. 
The free parameters of the high-precision NN-potentials are usually fixed by the experimental two-nucleon scattering data 
and describe the 2-body system extremely well\cite{Entem&Machleidt}. 
Unfortunately, these potentials produce unsatisfactory description of the 3- and 4-body systems\cite{Wiringa}.

A nuclear many-body system near equilibrium can be viewed as subject to a mean-field Harmonic Oscillator (HO) potential.
It is well-known that one can understand the magic numbers and the shell structure of nuclei within the 3-dimensional HO approximation plus a spin-orbit potential\cite{Haxel&Jensen}. 
Using the HO single-particle states one can write a general Hamiltonian with one- and two-body terms.
Despite the significant symmetry relations, 
due to rotational symmetry and due to the fermion exchange properties and the hermition requirement on the energy operator, 
the number of independent phenomenological parameters is often more than a dozen - usually it is of order of few hundred for the valence NN interaction alone. 
The independent parameters of the interaction are often fitted to experimental data by starting with some initial values that come from a relevant theory or model. 

Many of the high-precision NN-potentials, commonly used to build the microscopic interactions for multi-nucleon systems, 
have very complicated but methodically developed structure in terms of spin, iso-spin, and angular momentum components 
although sometimes there is a very complicated radial dependence. 
For example, the Argonne V18 potential has 18 different terms\cite{AV18}. 
Other potentials use non-local terms like CD-Bonn\cite{CD-Bonn} and Nijmegen\cite{Nijmegen}. 
However, when applied to A$>$2 systems all of these potentials have a serious difficulties 
that were eventually overcome by using three-body interactions\cite{TM'99, Wiringa, UIX}.

By the end of the twentieth century it become clear that a two-body interaction by itself is inadequate - even for the description of the lightest nuclei $2<A<5$. 
Comparative studies of various potentials, such as AV18, Nijmegen, CD-Bonn, and N$^3$LO, 
with or without three body terms have demonstrated the inadequacy of the pure two-body interactions and the need for three-body interaction terms\cite{Wiringa, Entem&Machleidt}. 
For example, all these interactions (AV18, Nijmegen, CD-Bonn, and N$^3$LO) describe very well the deuteron properties such as binding energy, radius, and quadruple moment 
but fail by more than 0.5 MeV to reproduce the binding energy of triton\cite{Entem&Machleidt} and underbind $^4$He by more than 4 MeV\cite{Wiringa}.

Although the meson-exchange approach was successful, it was clear that this phenomenological models should be derived from the underling QCD. 
Thus the ChPT approach became a prominent technique that produced the high-precision NN-potential N$^3$LO 
and then guided the researchers into the structure of the NNN- and NNNN-interactions\cite{Epelbaum, Entem&Machleidt, NNN_N2LO}.

\section{The NNN-body Interaction}

The use of the ChPT in the derivation of the nucleon interactions from QCD helped in identifying the mathematical form of various interaction terms along with the relevant parameters. Unfortunately, parameters related to contact terms in the interaction could not be determined. 
Thus, the strengths, $c_D$ and $c_E$, of the two-nucleon contact interaction with one-pion exchange to a third nucleon 
and the three-nucleon contact interaction are identified as undetermined parameters in the effective ChPT interaction\cite{Machleidt}. 
As such they need to be fixed by comparison with experiment.

\begin{figure}[htb]
\centering
\includegraphics[scale=0.34]{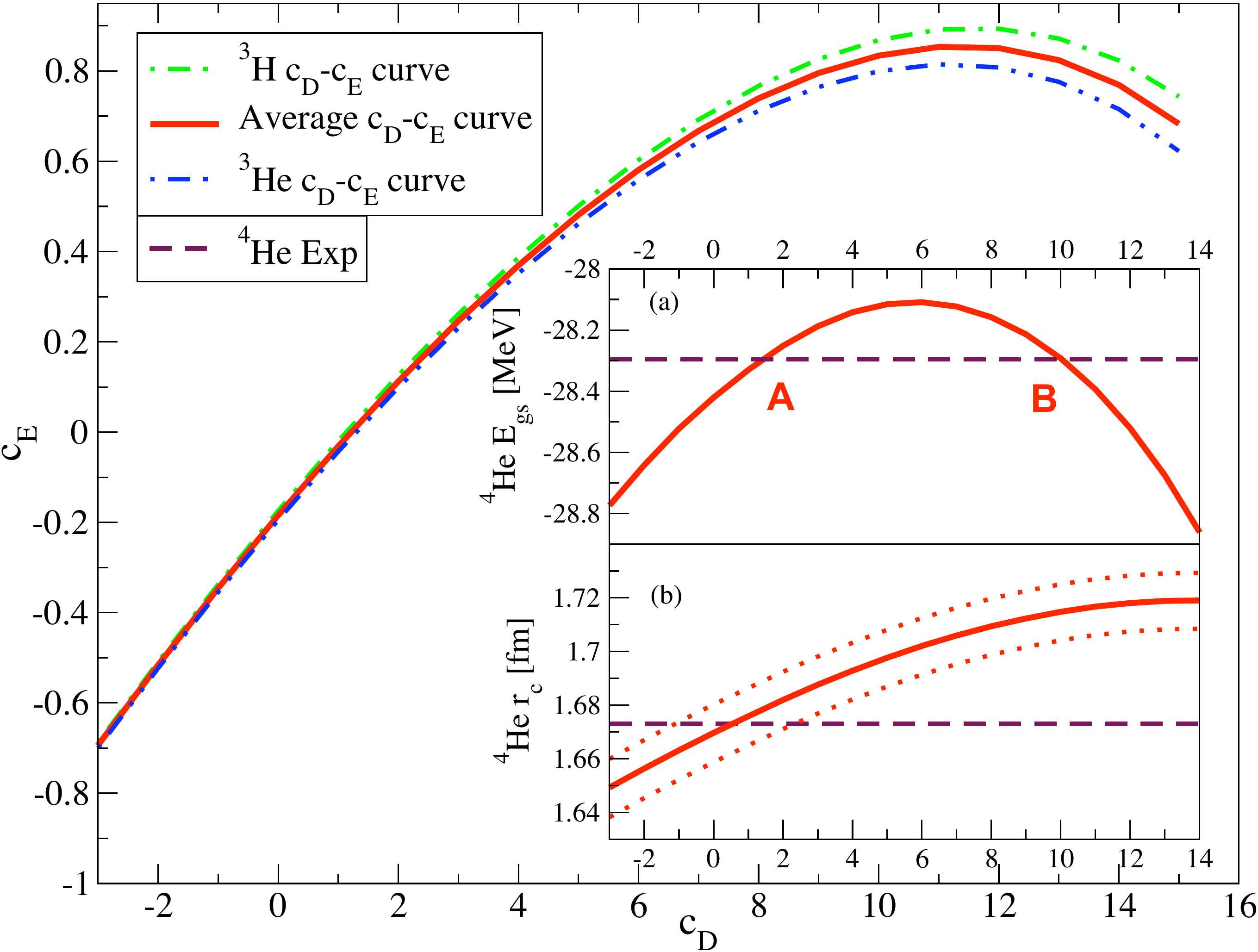}\\
\caption[]{Relations between $c_D$ and $c_E$ for which  the 
binding energy of $^3$H ($8.482$ MeV) and  $^3$He ($7.718$ MeV) are reproduced. 
(a) $^4$He ground-state energy along the averaged $c_D-c_E$ curve. 
The experimental $^4$He binding energy ($28.296$ MeV) is reproduced to within ~0.5 MeV over the entire range depicted.
(b) $^4$He charge radius $r_c$ along the averaged $c_D-c_E$ curve. Dotted lines represent 
the $r_c$ uncertainty due to the uncertainties in the proton charge radius.}
\label{CDCEcurve}
\end{figure}

The values of the parameters $c_D$ and $c_E$ that reproduce the binding energy of $^3$H and $^3$He within 0.5 keV of the experimental values correspond to two non-intersecting curves in the $c_D$ and $c_E$ plain as seen from Figure~\ref{CDCEcurve}. 
In order to further narrow down the range of $c_D$ values one considers the averaged $c_D-c_E$ curve 
and evaluates the binding energy of the $^4$He system that results in two possible physical regions denoted by A and B; 
where region A corresponds to $c_D$ of order $1$ and region B for $c_D$ of order $10$.
Finally, the charge radius of $^4$He points to the region A as the reasonable range of values for the $c_D$ parameter 
while the $c_E$ parameter is determined by the averaged $c_D-c_E$ curve\cite{Petr}.

Conceptually, there are three important concerns: 
First, the ChPT NN-potential was one order higher than the NNN-potential and no NNNN-potential was included. 
That is, the high-precision NN-potential was N$^3$LO (next-to-next-to-next-to-leading order)\cite{Entem&Machleidt} 
while the ChPT NNN-potential was at the N$^2$LO order\cite{NNN_N2LO} 
and the NNNN-potential\cite{Epelbaum} was not yet readily available. 
The second concern is that the range of the 3-body interaction parameter $c_D$ is determined by the properties of the 4-body system $^4$He; 
this, however, was resolved by a later study that used the $\beta^-$ decay of $^3$H into $^3$He 
and confirmed the physically relevant region A for the parameter $c_D$\cite{3Hto3He}. 
The third concern is related to the fact that these are high-precision studies and 
at this level of accuracy the difference between the proton and nucleon mass could be important for the A=3 systems\cite{Kamuntavicius'99}.

\section{Effective A-body Interactions}

In the previous section we discussed results obtained by using QCD derived interactions and the role of the NNN-interaction in the description of the light nuclei. 
Clearly 3- and 4-body interaction terms are predictions of the ChPT. 
Thus A-body interactions can be viewed as real physical interactions within the ChPT approach to nuclei. 
However, there is another way to arrive at A-body interactions that are phenomenological effective interactions 
since they are related to our inability to handle interacting systems in infinite Hilbert spaces\cite{AbodyHeff}. 
Since the quality of a model is judged by its ability to reproduce the experimental data, as far as computational models are concerned, 
an A-body interaction which gives results that agree well with the experimental data is also a physically relevant interaction.

In practice, we are computationally limited to a finite subspace of the infinite Hilbert space of the full quantum many-body problem. 
The subspace that we can access is defined by finite set of computationally convenient many-body basis states. 
For a suitable choice of basis we hope to have good overlaps with low-lying physical states of the system under study. 
If we imagine the exact solutions are available for analysis and apply a unitary transformation to those eigenstates, 
we can produce a transformed set of solutions maximally overlapping with our chosen basis space.

\begin{figure}[htb]
\centering
\includegraphics[scale=0.35]{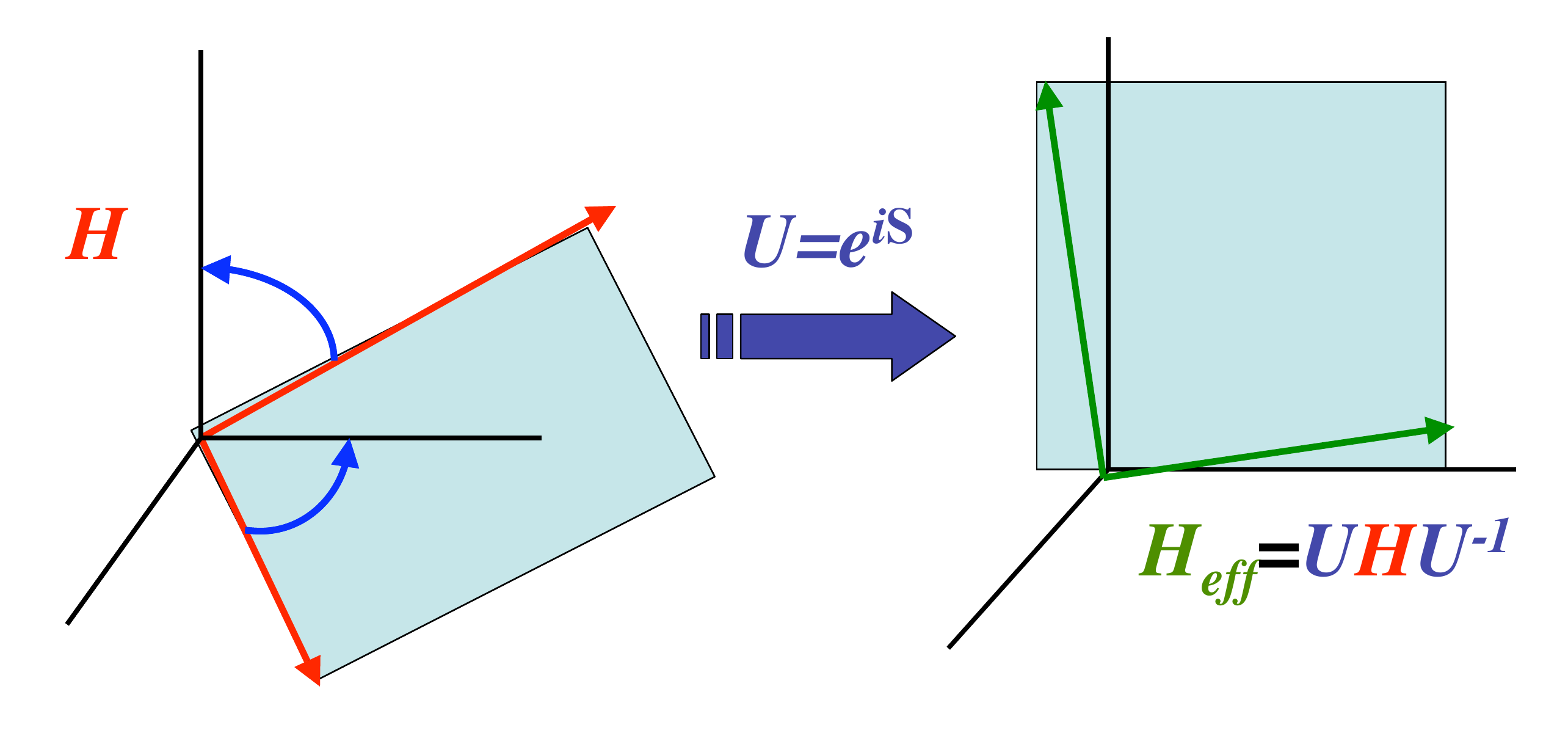}\\
\caption[]{Geometrical interpretation of the Okubo-Lee-Suzuki transformation method for construction of effective Hamiltonian operators.}
\label{Okubo-Lee-Suzuki}
\end{figure}

For example, one may be interested in the lowest two energy states of a system,  but would like to have some unitarily transformed version of these states that have maximal overlap with the two basis states that define a 2D computational space as shown in Figure~\ref{Okubo-Lee-Suzuki}. By finding the relevant unitary transformation U, one can define an effective Hamiltonian that would have the lowest two states within the 2D space as desired. Then this effective Hamiltonian could be used in the calculations of more complicated multi-particle systems, 
\textit{i.e.} one would find the unitarily transformed Hamiltonian that describes very well the low-energy states of a 2-body system in a mean field 
but within a Fock space that would be used later for an A-body system. 
Unfortunately, this transformation will turn any one- and two-body potential into a many-body effective interaction 
when applied to the relevant many-body system\cite{AbodyHeff}.

For A$>$4, it seems impractical at present to obtain the structure of the A-body interactions as derived from ChPT 
as it was previously done for the NNN- and the NNNN-interaction terms. 
Before embarking on the extensive undertaking required for including higher-body effective interactions, 
it would be very helpful to investigate a simple exactly solvable A-body interaction model that has few parameters and is applicable to real A-body systems.

\section{The Sn and Pb Isotopes}

At present, it seems impossible to be able to obtain the structure of the A-body interactions from ChPT for A$>$4. Therefore, in order to determine the relevance of the A-body interactions one should use the general form of an A-body interaction and then to try to determine some of the A-body interaction strengths. For this reason, one needs simple exactly solvable A-body interaction with few parameters that can be adjusted to the experimental data. Fortunately, there is such an interaction -\textit{ the Extended Pairing Interaction} (EPI)\cite{Feng}. This exactly solvable model is similar to the two-body proton-neutron pairing which was shown to be exactly solvable as well\cite{Dukelsky}. 

\begin{figure}[htbp]
\includegraphics[scale=0.47]{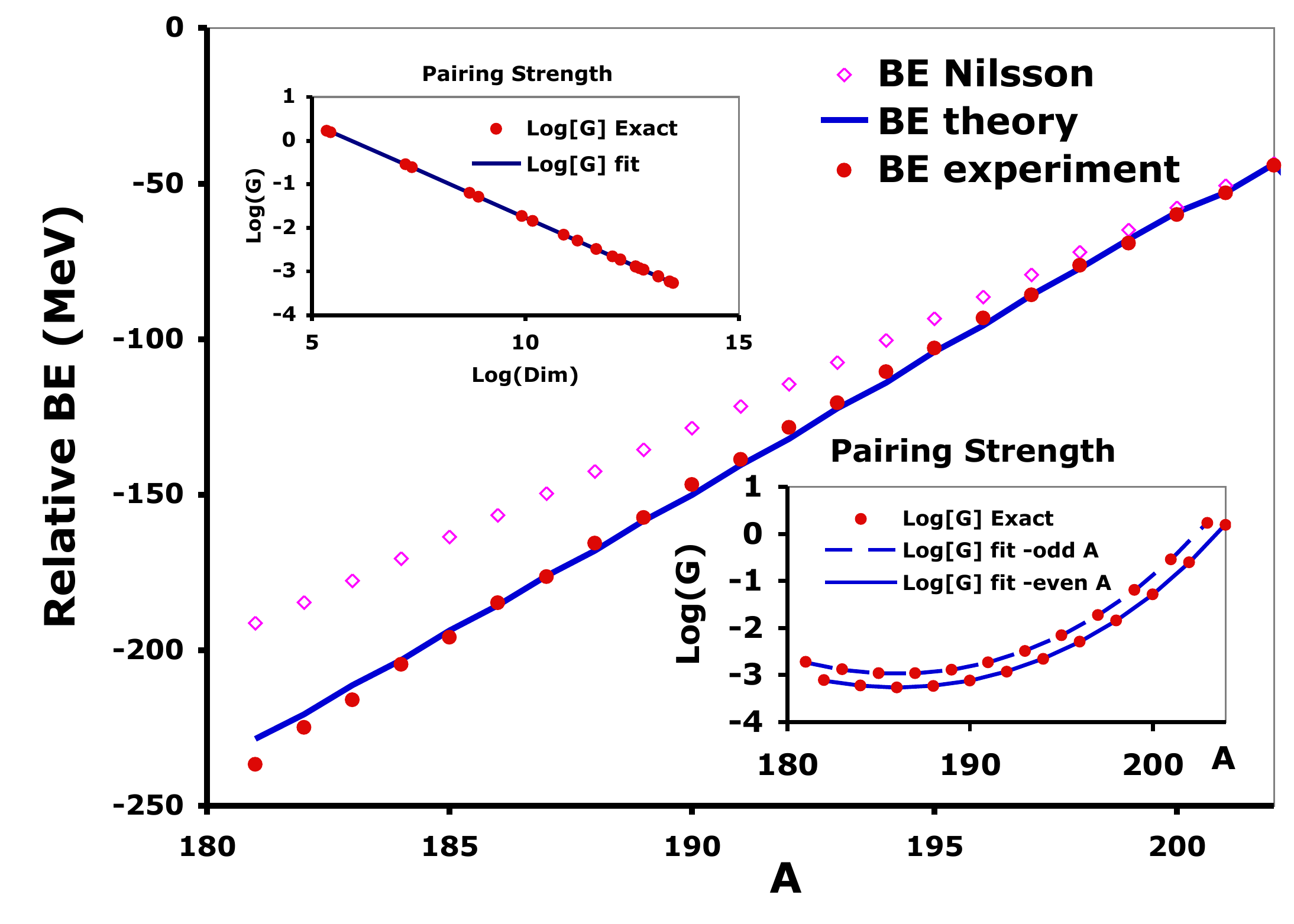}\\
%\centerline{\hbox{\epsfig{figure=Pb-isotopes_v1c.eps,height=2.6in,width=3.5in} }}
\caption{The solid line gives the theoretical RBE for the Pb isotopes relative to the $^{208}$Pb nucleus. 
The insets show the fit to the values of $G$ that reproduce exactly the experimental data using $^{164}$Pb core for the shell model. 
The lower inset shows the two fitting functions: 
$\log(G(A))=382.3502 - 4.1375 A + 0.0111 A^2$ for even values of $A$ and 
$\log(G(A))=391.6113 - 4.2374 A + 0.0114 A^2$ for odd values of $A$. 
The upper inset shows a fit to $G(A)$ that is inversely proportional to the size of the model space, ($\dim(A)$), 
that is valid not only for for even but also for odd values of $A$: 
$G(A)=366.7702\dim(A)^{-0.9972}$.
The Nilsson BE energy is the lowest energy of the non-interacting system.}
\label{Pb-isotopes}
\end{figure}

\begin{figure}[htb]
\centering
\includegraphics[scale=0.44]{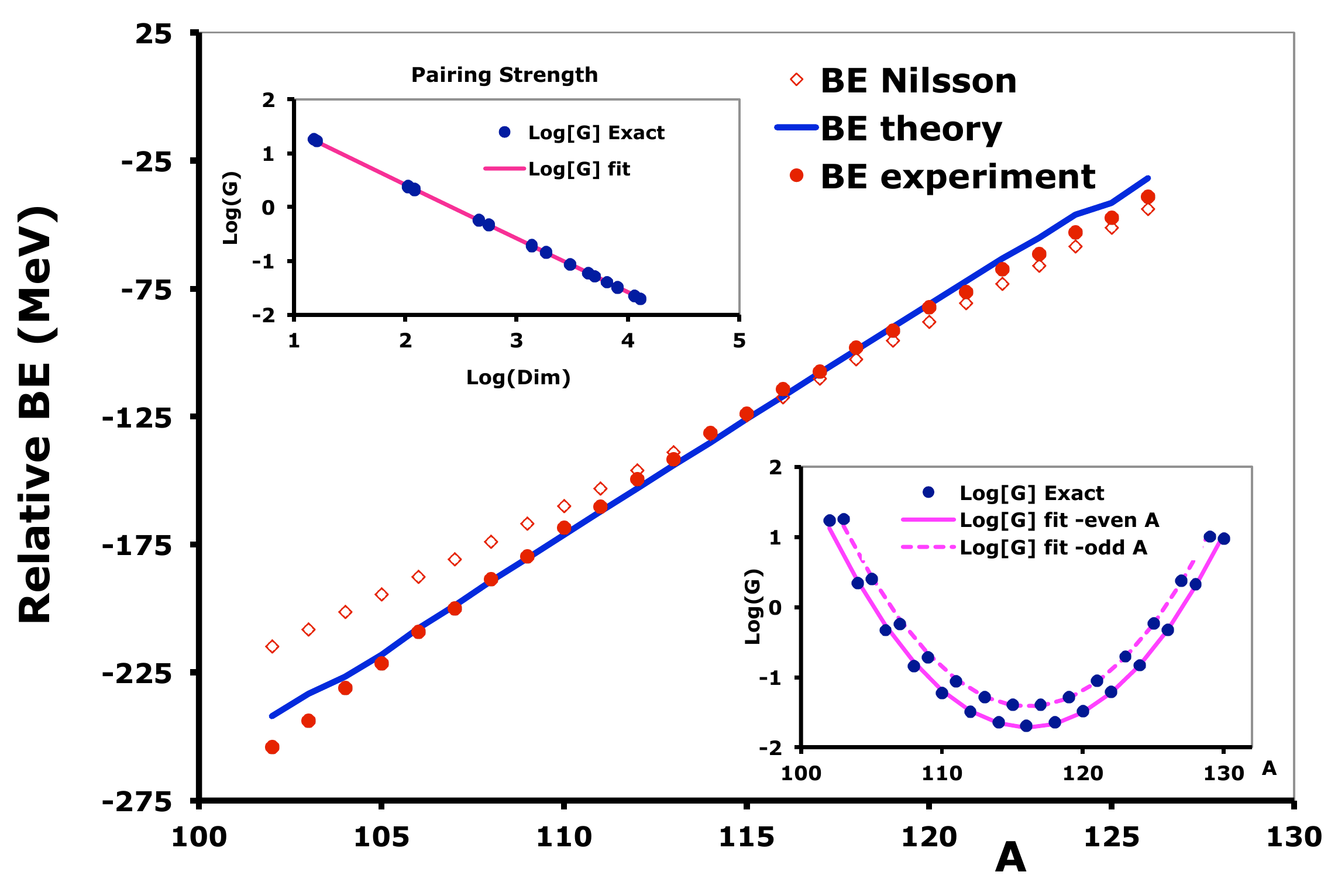}\\
\caption[]{The solid blue line gives the theoretical RBE for the Sn isotopes relative to the $^{132}$Sn nucleus. 
The insets show the fit to the values of $G$ that reproduce exactly the experimental data using $^{132}$Sn core. 
The lower inset shows the two second order polynomial fitting $\ln(G(A))=365.0584 - 6.4836 A + 0.0284A^2$ for even values of $A$ 
and $\ln(G(A))=398.2277 - 7.0349 A + 0.0307 A^2$ for odd values of $A$. 
The upper inset shows a fit to $G(A)$ that is inversely proportional to the size of the model space, ($\dim(A)$), that is valid
for even as well as odd values of $A$: $G(A)=\alpha\dim(A)^{-\beta}$ with $\alpha=259.436$ and $\beta=0.9985$.
The Nilsson BE energy is the lowest energy of the non-interacting system.}
\label{Sn_isotopes}
\end{figure}

Deformation is common in very heavy nuclei and this often justifies the success and application of the Nilsson model. Due to the space limitations many details and results of the current application of this exactly solvable Extended Pairing model are omitted, however, a more detailed paper is available \cite{VGG2004}. For the current application of the exactly solvable EPI model the single-particle energies are calculated using the Nilsson deformed shell model with parameters from \cite{Moller&Nix}. Experimental Binding Energies (BE) are taken from \cite{AudiG}. Theoretical Relative Binding Energies (RBE) are calculated relative to a specific nuclear system, $^{132}$Sn and $^{208}$Pb, for the cases considered. The RBE of the nucleus next to the core is used to determine an energy scale for the Nilsson single-particle energies. For an even number of neutrons, we considered only pairs of particles (hard bosons). For an odd number of neutrons, we apply Pauli blocking to the Fermi level of the last unpaired fermion and considered the remaining fermions as if they were an even fermion system.  The valence model space consists of the neutron single-particle levels between two closed shells with magic numbers 50-82 and 82-126. By using the exact solvability of the model, values of $G$ are determined so that the experimental and theoretical RBE match exactly as seen in Figure~\ref{Pb-isotopes} and Figure~\ref{Sn_isotopes}. The results are discussed in more details in Ref.\cite{VGG2004} and Ref.\cite{EPJ05}. 

Figure \ref{Pb-isotopes} shows results for the $^{181-202 }$Pb isotopes. The RBEs are relative to $^{208}$Pb which is set to zero, and the core nucleus is chosen to be $^{164}$Pb. Thus this calculations for the Pb-isotopes have a core nucleus $^{164}$Pb with a negative binding energy since the zero binding energy reference nucleus is set to be at $^{208}$Pb. One can see from Figure \ref{Pb-isotopes} that a quadratic fit to $\ln(G)$ as function of $A$ fits the data well. The fact that there is a correlation between the pairing strength $G$ and the size of the model space, reflected in the minimum of $G$ that is at the maximal model space dimension, prompted us to study $G(A)$ as function of the model space dimension $\dim(A)$. In this case the pairing strength $G(A)$ for all the 21 nuclei (A=$181-202$) was fit by a two parameter function $G(A)=\alpha /[\dim(A)]^{\beta}$ with the values of the parameters taken to be $\alpha=366.7702$ and $\beta=0.9972$. Similar results have been obtained for the Sn-isotopes as well by using $^{132}$Sn as zero RBE system (see Figure~\ref{Sn_isotopes}).

The Sn isotope chain is unique in the sense that there are two doubly magic members, the $^{100}$Sn and $^{132}$Sn.
This allows us to use $^{132}$Sn as zero RBE system with holes as it has been done for the Pb case\cite{EPJ05}.
Again, there is a simple expression that works for even and odd systems simultaneously: $G(A)=\alpha\dim(A)^{-\beta}$; 
with $\alpha=259.436$ and $\beta=0.9985$ for $^{132}$Sn as reference RBE system. 
Having a one parameter expression ( one practically has $\beta=1$) for the extended pairing strength 
in such long chains of isotopes ($\sim 20$ nuclei in the chain) is remarkable!

In the light of the above, it seems that the next step should be a study that tracks the results as a function of the increasing size of the model space to confirm or refute the $\log-\log$ relation. Such a study could also address other questions such as the effect of the core binding energy as a function of the deformation that is used in the Nilsson model to derive the single-particle energies. Using a Woods-Saxon potential or other methods to generate more realistic single-particle energies is another opportunity for further studies.

\section{Conclusion}
In conclusion, we studied relative binding energies of nuclei in two isotopic chains, $^{100-130}$Sn and $^{181-202 }$Pb, using the A-body Extended Pairing Interaction \cite{Feng} by using Nilsson single-particle energies as the input mean-field energies. Overall, the results suggest that the model is applicable to neighboring heavy nuclei and provides, within a pure shell-model approach, an alternative mean of calculating relative binding energies. In order to achieve that, the pairing strength is allowed to change as a smooth function of the model space dimension. It is important to understand that the A-dependence of G is indirect, since G only depends on the model space dimension, which by itself is different for different nuclei. In particular, in all the cases studied $\ln(G)$ has a smooth quadratic behavior for even and odd $A$ with a minimum in the middle of the model space where the dimensionality of the space is a maximal; $\ln(G)$ for even $A$ and odd $A$ are very similar which suggests that further detailed analyses may result in the same functional form for even $A$ and odd $A$ isotopes as found in the case of the Pb-isotopes and Sn-isotopes. It is a non-trivial result that $G$ is inversely proportional to the space dimension $\dim$ in the two cases found (Pb-isotopes and Sn-isotopes) which requires further studies. 

In this paper we have presented evidence for the need to better understanding of the NNN-, NNNN-, and A-body interactions in nuclei 
either derived from ChPT or from a phenomenological considerations. Therefore, one has to build A-body computational technology in the next generations of nuclear modeling codes.While the motivation for considering A-body interaction in the light-nuclei is strong as based on the ChPT QCD derived interactions, one is left to wonder if A-body interactions are also relevant to heavy nuclei. The results obtained with the help of the Extended Pairing Interaction, in particular the Sn and Pb isotopes discussed here seem to confirm the idea that A-body interactions are needed to understand better the binding energy of heavy nuclei. Often the imagination cannot capture all the possible implications and uses of an exactly solvable model.  Beside the current applications of the Extended Pairing Interaction, one can also see that it would be a useful verification tool for A-body computational codes as well. 

\section*{Acknowledgements}
The author is particularly grateful to P. Navr\'atil, J. P. Vary, J. P. Draayer, F. Pan, and many more of his colleagues and collaborators for their support and research opportunities over the years.

%\bibliographystyle{unsrt}
%\bibliography{B2BIPinNS}
%\end{document}

\end{document}